\newcommand{\beq}{\begin{quote}}
\newcommand{\enq}{\end{quote}}
\newcommand{\be}{\begin{equation}}
\newcommand{\en}{\end{equation}}

\documentstyle[12pt,epsfig]{article}
\begin{document}
\title{  Hooke's Memorandum on the
development of orbital dynamics
} 
\date{}
\author{ 
}
\maketitle
\begin{abstract}
I discuss  a  memorandum entitled 
'' A True state of the Case
and Controversy between $Sir$ Isaak Newton \& $Dr$ Robert Hooke as
the Priority of that Noble Hypothesis of Motion of ye Planets about
ye Sun as their Centers'',  where Robert  Hooke summarized  his
seminal contributions to  the physics of orbital motion
and to the  theory of universal gravitation.

\end{abstract}

In a brief handwritten but undated  memorandum entitled 
'' A True state of the Case
and Controversy between $S^r$ Isaak Newton \& $D^r$ Robert Hooke as
the Priority of that Noble Hypothesis of Motion of ye Planets about 
ye Sun as their Centers'' \cite{memo1}, Hooke recounted his 
{\it hypothesis} for the physics of orbital motion 
and his theory of universal gravitation. Hooke's memorandum, 
which remained unpublished during his lifetime, is historically quite accurate,
contradicting numerous criticisms of his contemporaries and  
historians of science that Hooke always claimed for himself  more credit than 
he actually deserved. In fact, to support his ''priority'' 
Hooke quoted verbatim from several extant documents: 
the transcript of his lecture on {\it Planetary
Movements as a Mechanical Problem} given at the Royal 
Society on May 23, 1666 \cite{transact}, his short
(28 pages) monograph, {\it  An Attempt to prove 
the motion of the Earth by Observations} published in 1674 \cite{mono1}, 
and  his lengthy correspondence in the Fall of 1679  with Isaac Newton 
\cite{turnbull}. However,  Hooke did not mention
his remarkable geometrical implementation of orbital motion 
for central force motion, see Fig. 1, 
based on the application of his physical principles, which was 
found only recently in a manuscript dated
Sept. 1685 \cite{pugliese} \cite{michaelh}. Unfortunately, Hooke did 
not publish this manuscript and related work 
in spite of Edmond Halley's urging him 
''... that unless he produce another differing demonstration [from Newton's],
and let the world judge of it, neither I nor any one else can believe
it'' \cite{halley0}. 
\begin{figure}
\begin{center}
\epsfxsize=\columnwidth
\epsfig{file=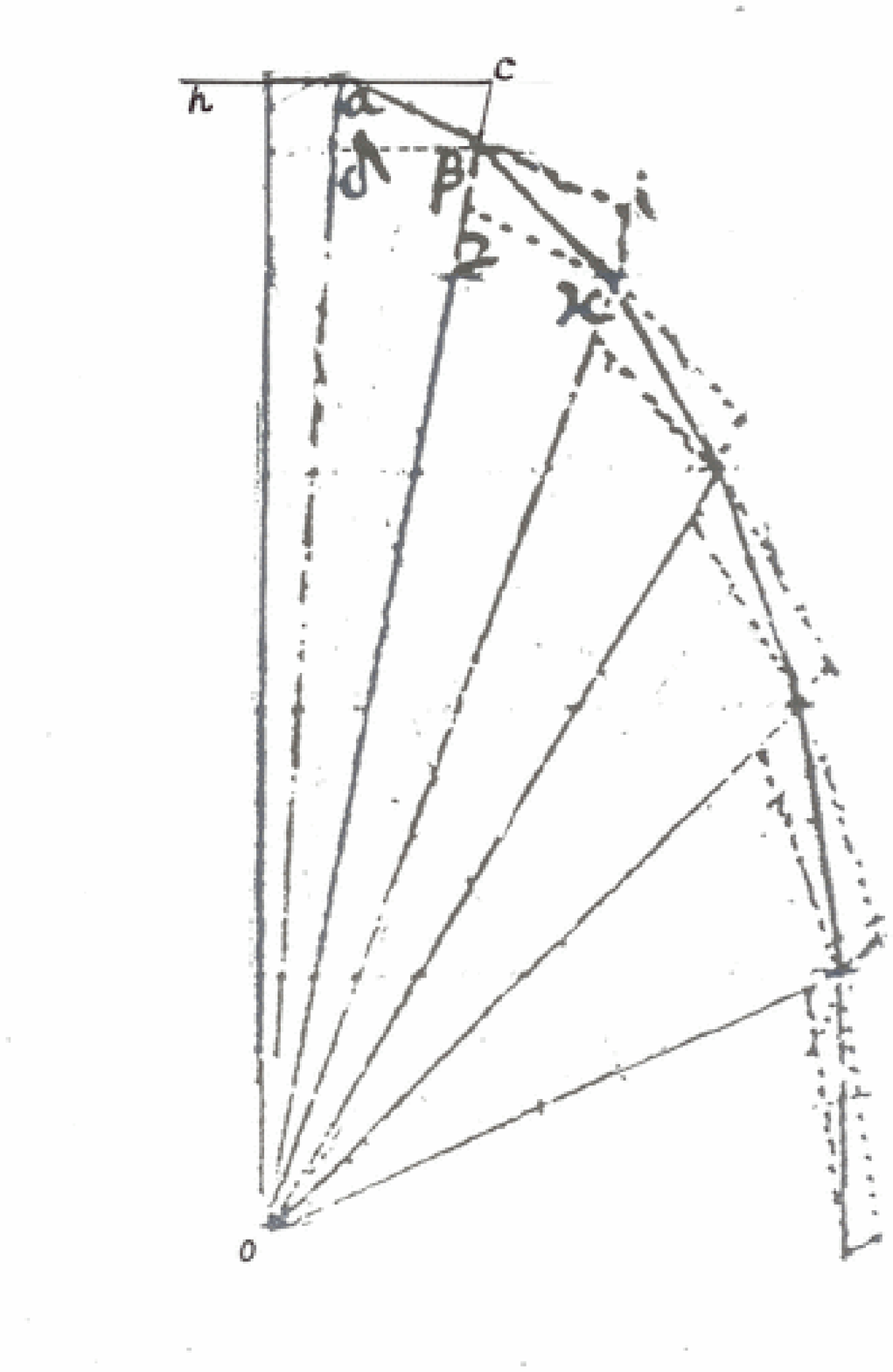, width=10cm}
\end{center}
\caption{
The upper right hand
part of Hooke's Sept. 1685 diagram,
with some auxiliary lines deleted, showing his geometrical
construction for a discrete approximation to an elliptic orbit 
rotating clockwise under the action of a sequence of radial 
impulses  which vary linearly with the
distance from the center at $O$. For details, see ref. \cite{michaelh}
}
\label{Fig. 1}
\end{figure}
It can be seen that Hooke's geometical construction is virtually the 
same as the one described 
by Newton, see Fig. 2, in connection with his proof of Kepler's area law 
in {\it De Motu}, a short draft that Newton sent to the Royal Society in 1684, 
which subsequently he  expanded into his monumental work, 
the {\it Principia}\cite{cohen1}.

\begin{figure}
\begin{center}
\epsfxsize=\columnwidth
\epsfig{file=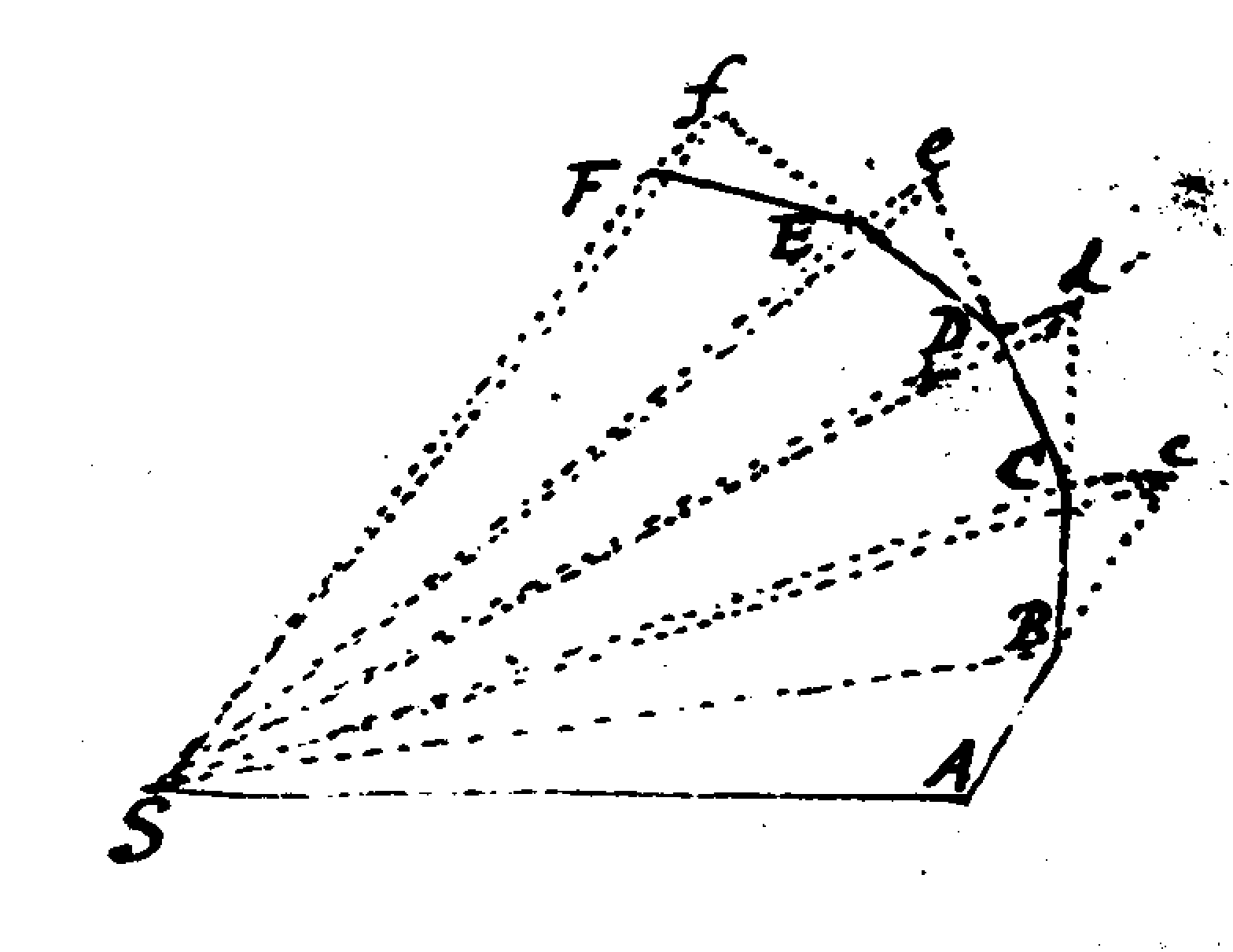, width=10cm}
\end{center}
\caption{ Diagram in {\it De Motu} associated with Newton's proof
of Kepler's area law, showing the construction of  a discrete orbit 
rotating counterclockwise under the action of a sequence of radial 
impulses of unspecified magnitude with  center at $S$.
}
\label{Fig. 2}
\end{figure}

In his memorandum, Hooke recounts that already in 1666 he 
had suggested that the motion of planets around the sun 
can be understood by  the ''inflection of a direct motion 
[inertial motion] into a curve by a supervening attractive principle''
\cite{memo1}, the gravitational attraction of the sun. 
He supported this novel physical insight by a mechanical 
{\it analog}, namely, the motion of a  conical
pendulum, which he demonstrated experimentally by hanging a weight
from the ceiling of the room where he lectured to members
of the Royal Society. He also analyzed the pendulum  motion 
mathematically, showing that the component of the net
force directed towards axis of the pendulum increased 
linearly with the distance, recognizing that it `` seems
to be otherwise in the attraction of the sun...''   \cite{transact}.
In his 1674 monograph, which contains his first Cutlerian lecture 
at Gresham College given in 1670,  Hooke restated 
his  physical principles for the origin of  curved motion by the action
of an  attractive force, and then enunciated his 
``supposition'' for the law of  universal gravitation:  that  '' all celestiall bodys 
whatsoever have an attraction or a gravitating power towards their own Centers, 
whereby they attract not only their own parts, \& keep them from flying
from them, as we may observed the Earth to do, but that they do also attract
all the other Celestiall Bodies which are within the sphere of their activity''
\cite{memo1} \cite{mono1} 
As far as we know, this statement is the first  {\it published} suggestion that the   
gravitational force which attracts objects to the surface of the earth 
also acts between  celestial bodies.  Hooke elaborated this 
theory  by supposing that ''...  not only the Sun and Moon
have an influence upon the body  and  motion of the Earth 
and the Earth upon them, but that Mercury, also Venus, Mars, Saturn and Jupiter
by their attractive powers, have considerable influence upon its motion
as in the same manner the corresponding attractive power of the Earth
hath a considerable influence upon  every one of their
motions also ''. The great novelty of this extension of terrestial gravity
to celestial bodies  is underscored by the disbelief with which it
was received by some of the greatest scientific minds  in Europe when it was  
proposed again, in similar form, about 13 years later
by Newton  in  Book 3 of  his {\it Principia}   
\footnote { This book is  entitled {\it The System of the
World} which are the  words that  Hooke used to  introduce  his theory of
universal gravitation in his 1674 tract {\it An Attempt to prove
the Motion of the Earth by Observations}}. 
After first reading the {\it Principia}, Christiaan Huygens admitted that 
it never occured 
to him  ''to extend the action of gravity
to such great distances as those between the sun and the planets or between
the moon and the earth'' \cite {cohen155},  while
Gotfried Leibniz declared that ''the introduction of gravitation
of matter towards matter  is in effect to return to occult qualities and,
even worse, to inexplicable ones \cite{leibniz}.
Actually, an account of Hooke's 1674  monograph introducing the
idea of universal gravitation had appeared in {\it The Philosophical
Transactiona}, Vol IX, {\bf 101}, 12, (1674), and four issues later
there appeared extracts of several letters containing  comments 
including one by Huygens. Evidently, after the publication
of the {\it Principia} in 1687,  Hooke's priority in proposing
universal gravitation had been forgotten. 

In the first edition of the {\it Principia}  
Hooke's  early proposal for universal 
gravitation was not mentioned, while in the second edition (1713), 
Newton left it to his editor, Roger Cotes, 
to admit in an editor's preface, 
'' that the force of gravity is in all bodies universally 
others have suspected or imagined, but 
Newton was the first and only one who was able to demonstrate 
it from phenomena and to make it a solid
foundation  for his brilliant theories''.  
Even this small concesion to ''others'',
was left out in Newton's  third and final edition (1726) 
of the {\it Principia}.  
Apparently, after hearing of Hooke's priority complains, 
Newton eliminated many references to Hooke
in earlier drafts of his {\it Principia}. In a letter to Halley,
Newton complained
that ''... he [Hooke] knew not how to go about it. Now is not this
very fine? Mathematicians that find out, settle \& and do all the businness
must content themselves with being nothing but dry calculators \& drudges \& and
another that does nothing but pretend \& grasp at all things must carry
away all the invention as well as those who were to follow him as of those 
that went before him '' \cite{halley2}. Probably  prompted
by Newton,  Cotes added the  remark that 
''I can hear some people disagreeing with this conclusion[about universal
gravitation] and muttering something or other about occult qualities. 
They are always
prattling on and on to the effect that gravity is something occult, and
that occult causes are to be banished completely from philosophy.''
\cite{cotes1}. But this remark applies only to Huygens and Leibniz, 
and not to Hooke who nowhere in his writings 
considered gravity to be an {\it occult} quantitiy. 

In his memorandum , Hooke did  not claim to know 
how the  gravitational force varies  with  distance,   
supposing only '' that these attractive powers are so much the more powerful
in operating, by how much the nearer the body wrought upon is to their
own Center''. Instead, Hooke proposed that this dependence  be determined
experimentally,  and predicted that it ''will mightily assist the Astronomer to 
reduce all Celestiall Motions to a certain rule, which I doubt will never 
be done true without it'' \cite{mono1}. 
Finally, Hooke recalled that  he communicated 
his principles for  orbital motion in a correspondence 
with Newton. In a letter dated  Nov. 24, 1679, he explicitly
asked Newton ''as a great favour if you shall please to communicate your
objections against my Hypothesis or opinion of mine particularly if you
will let me know your thoughts of that compounding of celestiall motions
of the planets of a direct motion by the tangent \& an attractive motion 
towards the centrall body'' \cite{mono1} \cite{turnbull}.
 Hooke notes that ''in answer to this Newton
pretends he knew not Hooke's Hypoth. ''\footnote{ this comment also appears
as an  insertion  Hooke made in his copy of Newton's letter.}, 
referring to Newton's response
on Nov. 28, 1679
that  ''...perhaps you will incline the more to believe me when I tell
you that I did not before the receipt of your last letter, so much as
heare (that I remember) of your Hypothesis of compounding the celestial
motions of the Planets, of a direct motion by a tangent to the curve... ''
\cite{mono1} \cite{turnbull} 
But in the same letter Newton also remarked  that '' I am glad to heare that so
considerable a discovery you have made of the earth's annual parallax
is seconded by Mr. Flamstead's Observations''. Since Hooke 
had not mentioned  his own role in this supposed discovery, 
this remark indicates that Newton already was familiar 
with Hooke's 1674  monograph, where Hooke had published his
observations which he incorrectly had interpreted as due 
to  the  earth's annual parallax.  
But in this monograph Hooke  also enunciated his principles of 
orbital dynamics, which Newton '' pretended'' not to have heard.

The question still remains what, if anything, did Newton learn from his 
1679 correspondence with Hooke?
Newton's early notebook, the {\it Waste book}, indicates that
by 1664  he was already  analysing  uniform  circular motion  
by the action of a sequence of impulses on a moving body 
that are directed  towards the center of the circular orbit \cite{herivel}. 
Therefore, it is incorrect to assert, as several historians of science 
have done, that Newton had  learned this  approach 
to orbital motion from Hooke \cite{westfall}.
But it is surprising that in his Nov. 28 letter to Hooke, 
Newton claimed  that he was unaware that Hooke had had similar views
on orbital motion, because Newton had read Hooke's 1674 monograph.  
In his memorandum, Hooke recalled  that in his  response letter  
he had reminded  Newton that  `` I could add many other considerations which are
consonant with my Theory of Circular motions compounded by a Direct motion
and an attractive one to a center''\cite{memo1}, \cite{p304}.
Later on, in his 1686  correspondence with Halley
regarding what he had heard about
Hooke's priority claims, Newton focused  mainly on 
the discovery of the inverse square 
dependence of  the gravitational force, neglecting to mention 
Hooke's earlier formulation of the principles of orbital dynamics and 
the theory of universal  gravitation.

According to David Gregory, who visited Newton at Cambridge in 1694,
'' I saw a manuscript [written] before 1669 ... where all the foundations of his
philosophy are laid down: namely the gravity of the Moon to the Earth, and of the
planets to the Sun. And in fact all these even then are subject to calculation...''
\cite{herivel1}. This manuscript, which is found among Newton's still
exisiting  papers, indicates that
by 1669 Newton had gone further than Hooke, having rediscovered the 
mathematical relation for the  
radial acceleration or central force in the case of  uniform circular motion
that had been found earlier by  Huygens, but
not published  by him until 1673, when his {\it Horologium Oscillatorium} 
first appeared.  Newton  applied this relation to planetary motion, 
and by assuming that
it satisfied Kepler's harmonic law, he found that '' the endevaours
of receding from the Sun will be reciprocally as the squares of the
distance from the Sun'' \cite{herivel1}.
Newton  assumed that such a  dependence
on distance applied  also to the  force  attracting the  moon  
to the earth,  which  he attempted to identify with the gravitational
force acting on bodies at the surface of the earth.
But due to an error in the value for the radius 
of the earth which he used in his calculations, 
he was misled into thinking  that an  inverse square dependence
was not accurate for terrestial gravity.
Actually, it was not until after he discovered his error around 1685,
by applying Piccard's correct value for the earth's radius to his
earlier calculation, that he  proved,  by a remarkable 
mathematical feat that this dependence is valid up to the surface
of any spherical body.  \footnote{ In his book,
{\it Newton's Principia for the Common reader}, Chandrasehkar  calls 
it one of Newton's  ''superb theorems''.}
Without this proof, however, the moon test
for  the universality of gravity is not possible, although 
usually this is forgotten. In his 1686  correspondence with Halley, in
which he rejected accusations that he had learned about the inverse square
dependence from Hooke, Newton  remarked that '' Mr. Hook without knowing
what I have found since his letters to me, can know no more than the
proportion was duplicate {\it quam proxim\'e} [approximately] 
at great distances from the
center, \& only guessed it to be so accurately, \&  guess amiss in
extending that proportion down to the very center...'' . Newton repeatedly
used the word {\it guess} to indicate that Hooke had not provided any 
mathematical proof
for his supposition ''that the Attraction  always is in a duplicate
proportion to the Distance  fromt he Center Reciprocally'' as Hooke had written
to him \cite{p309}. In a letter to
Halley, Newton pointed out that in this ''Theory I am plainly before
Mr Hook. For he, about a year after [1673], in his 
{\it Attempt to prove the Motion of the Earth}, declared expressely 
that the degrees by which gravity decreased
he had not then experimentally verified, that is he knew not how to gather
it from phenomena, \& therefore  he there recomends it to the prosecution
of others'' \cite{p446}. Newton also asserted  that 
Hooke had extended the inverse square proportion to the interiour 
of the earth. But, instead, Hooke had correctly pointed 
that inside the earth the gravitational force varies linearly with 
the distance from the center, stating that  '' I rather Conceive 
that the more the body approaches the Center, the lesse will it be 
Urged by the attraction- possibly somewhat like the Gravitation 
on a pendulum or a body moved in a
Concave Sphere where the power Continually Decrease the neerer the body 
inclines to a horizontal motion...'' \cite{p309}.

On  Dec. 13, 1679 Newton wrote a remarkable letter to Hooke \cite{p307},  
which demonstrates
that by that time Newton had gained a deep understanding of the physics  of 
central force motion, and provides evidence that he
had  developed a very good  approximate mathematical 
method to calculate the orbits  
for various central forces \cite{michael}. The letter includes a diagram,
Fig. 3, which shows the trajectory of a body moving under the action of
a central force which has a constant magnitude. 
\begin{figure}
\begin{center}
\epsfxsize=\columnwidth
\epsfig{file=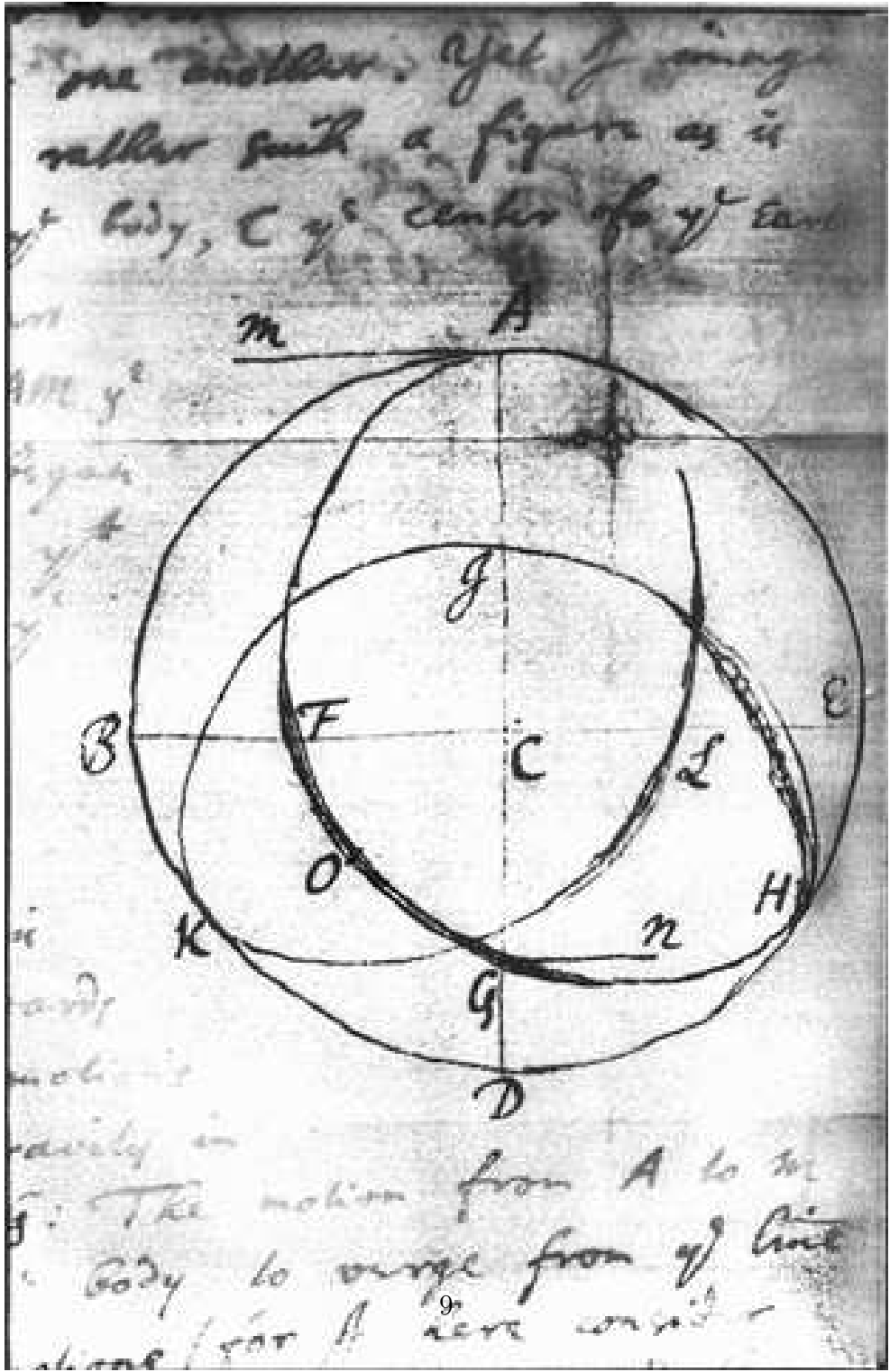, width=13cm}
\end{center}
\caption{ Diagram in Newton's Dec. 13, 1679 letter
to Hooke, showing a curve $AFOGHIKL$ for the  approximate orbit 
of a body moving under the action of a constant central force.
}
\label{Fig. 3}
\end{figure}
Hooke promptly responded
that ''Your Calculation of the Curve by a body attracted
by an equall power at all Distances from the center Such as that of
a ball Rouling in an inverted Concave Cone is right and the two auges
[farthest points from the center of force] will not unite by about
a third of a revolution'' \cite{hooke3}. Hooke must have been astounded that
Newton could calculate a trajectory which previously he had
observed in one of his mechanical experiments to understand orbital motion.
In Fig. 4  I show  a stroboscopic photograph of a steel ball
rolling inside an inverted cone which closely resembles the trajectory
shown in Newton's diagram, Fig. 3, and attests to  Hooke's 
careful observation. 
\begin{figure}
\begin{center}
\epsfxsize=\columnwidth
\epsfig{file=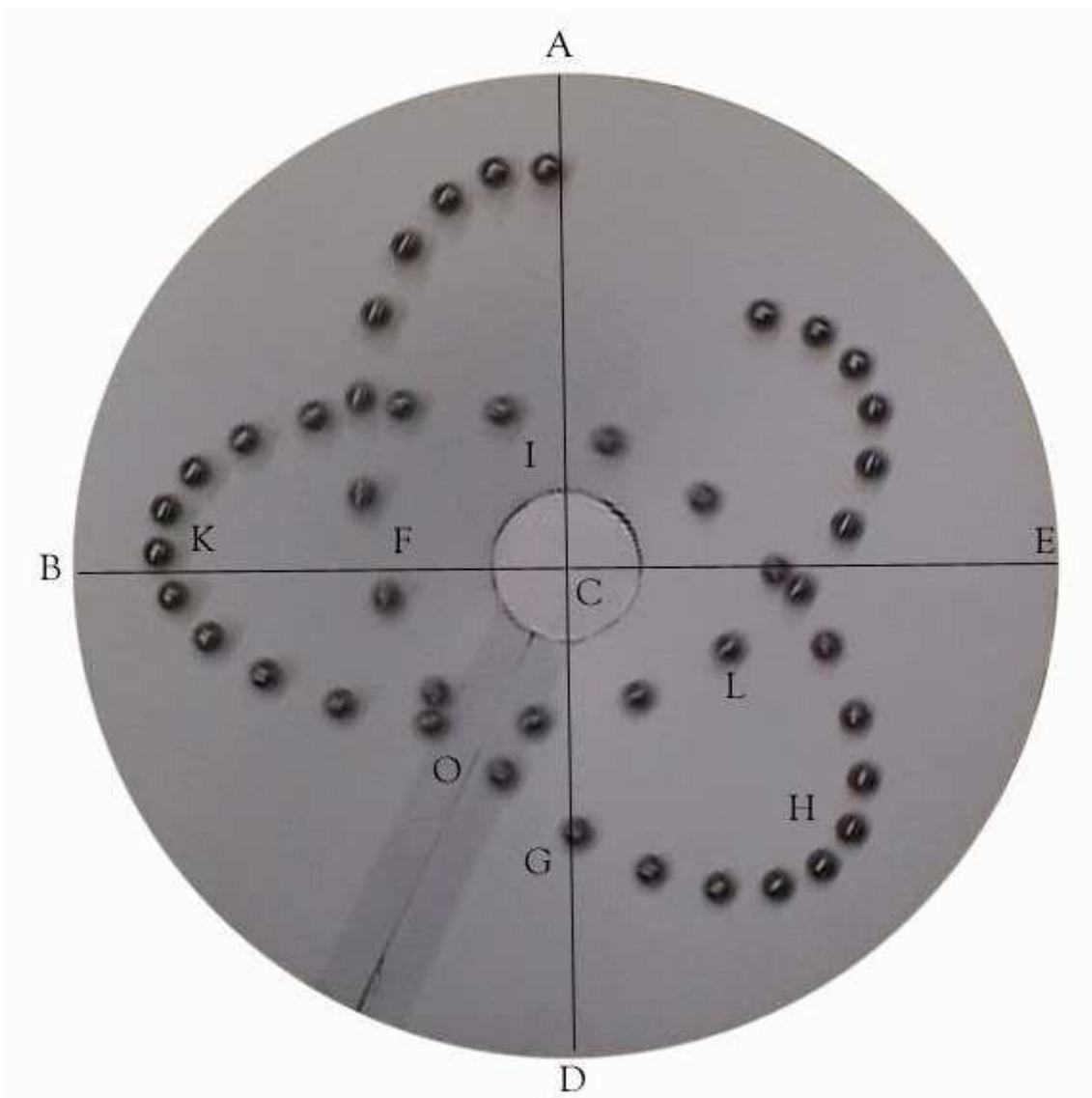, width=15cm}
\end{center}
\caption{ Stroboscopic photograph reproducing Hooke's observation
of   '' a ball Roulling in an inverted Concave Cone''.
}
\label{Fig. 4}
\end{figure}
In the text to his letter to Hooke, Newton also
discussed the changes in the orbit when the force 
increases with decreasing distance to the center, or in his words,
'' Thus I conceive it
would be if gravity where the same at all distances from the center. But
if it be supposed greater nearer the center the point $O$ [nearest to
the center C] may fall in the
line $CD$ or in the angle $DCE$  or in other angles that follow, or even
nowhere. For the increase of gravity in the descent may be supposed such 
that the body shall by an infinite number of spiral revolutions descend
continually till it cross the center by motion transcendentally swift''
Although in his letter to Hooke, Newton did not identify the force law
that would  lead to such an extraordinary orbit with ''an infinite number
of spiral revolutions'', later in 1684 he amplified his  description 
in a Scholium to an early draft of the {\it Principia} \cite{whiteside2}, 
and he  revealed  that this force 
depends inversely as the cube power of the radial distance. 
This Scholium, however, was not included in the final draft
of the {\it Principia},  and it has been generally 
ignored in the past. Thus,  
it is evident that by the time of his correspondence with 
Hooke, Newton already had developed a fairly sophisticated 
method to calculate orbital motion for central forces.
I have given arguments \cite{michael} that  Newton's  method was based on 
his observation that for central forces  the component of the force 
normal to the orbit  determines  its radius of curvature
by the Huygens-Newton relation for circular motion, 
provided that the velocity is known.  
Newton indicated this connection in a cryptic remark in his 1664 notebook:
'' If the body b moves in an Ellipsis, then its force in each point
(if its motion [velocity] be given) may be found by the tangent circle of equal
crookedness [curvature] with that point of the Ellipsis''\cite{herivel}.
But in this curvature approach  it is difficult to see that 
Kepler's area law [conservation of angular momentum] is a consequence
of the action of  central forces \cite{michael}. 
Newton discovered this fundamental connection,
which became a cornerstone of his {\it Principia} as Proposition 1 in Book 1,
only after his correspondence with Hooke. To prove this theorem
Newton first had to  discretize the continuous central force
by a series of impulses, and then apply to general orbital 
motion the principles advocated for a long time by Hooke,  
which he had explained to Newton  in his 1679 correspondence as 
'' compounding the direct 
motion with an inflection towards the center of force''. Although previously
Newton had applied such a  decomposition to  uniform circular motion, evidently 
the impetus for considering it for general motion came from Hooke, yet
Newton vehemently denied that he learned anything from him, admitting only
that '' ...his correcting my Spiral occasioned my finding the Theorem
by which I afterward examined the Ellipsis; yet I am not beholden to him
for any light into that business but only for the diversion he gave me from
my other studies  and for his domaticalness in writing as if he had found
the motion  in the Ellipsis, which inclined me to try it after I saw by what
method it was to be done...'' \cite{p446}. 
But  without Hooke's crucial intervention in 1679, it is most likely that
Newton would have continued with his ''other studies'' [alchemy and theology].
Then, when in  the Fall of 1684 Halley was travelling 
back to London after burying his father in Lincolnshire \cite{whiteside3},
and decided to visit Newton in Cambridge, he would not have been prepared to respond '' an ellipsis" 
to Halley's famous  question \footnote{according to Newton's later recollection,
as told by Abraham De Moivre}
''what he thought the Curve would be that would be described by the
Planets supposing the  attraction towards the Sun to be reciprocal to
the square of their  distance from it''\cite {westfall2}.


\begin{thebibliography}{30}

\bibitem{memo1}  This is an undated manuscript in Hooke's collection at the
                  Trinity Library which is reproduced in  R.T. Gunther,
                  {\it Early Science in Oxford}, Vol. X (Oxford University, 
                     Oxford, 1930) pp. 57-60

\bibitem{transact} T. Birch, {\it The History of the Royal Society of London}
                    (Royal Society, London, 1756-57), pp. 91-912; 
                    Gunther, ref. \cite{memo1}, Vol. VI, p. 256.

\bibitem{mono1}  R. Hooke, {\it An Attempt to Prove the Motion of the Earth
                    from Observations}, (London,1674), reproduced  in Gunther, 
                    ref. 1, Vol VII, pp. 1-28.

\bibitem{turnbull} {\it The Correspondence of Isaac Newton}, Vol. II, 
                   1676-1687, edited by H.W. Turnbull (Cambridge University Press,
                   1960)

\bibitem{halley0}   Letter from Halley to Newton dated June 29, 1686, {\it
                    The Correspondence of Isaac Newton} ref. \cite{turnbull}
                    p.442

\bibitem{pugliese}  P.J. Pugliese, {\it Robert Hooke and the Dynamics of Motion
                    in a Curved Path}, in M. Hunter and S. Schaffer, ''Robert Hooke,
                    New Studies'' (Boydell, Woodbridge, 1989) pp. 181-205

\bibitem{michaelh}  M. Nauenberg, {\it Hooke, Orbital Notion, and Newton's 
                    Principia}, American Journal of Physics {\bf 62} (1994) 331-350;
                    M. Nauenberg, {On Hooke's 1685 Manuscript on Orbital Mechanics}
                    Letter to the editor,  Historia Mathematica {\bf 25} (1998), 89-93
                    M. Nauenberg, {\it Robert Hooke's Seminal Contributions to
                    Orbital Dynamics},  Physics in Perspective, March 2005, ??

\bibitem{cohen1}    I. Newton, {\it Mathematical Principles of Natural Philosophy},
                    A new translation by I.B. Cohen and Anne Whitmann, with
                    a guide to Newton's {\it Principa} by I. B. Cohen (University
                    of California Press, Berkeley 1999) 

\bibitem{cohen155}  Quoted in ref. \cite{cohen1}, p. 155

\bibitem{leibniz}   Quoted in ref. \cite{cohen1}, p. 153

\bibitem{halley2}   Letter from Newton to Halley, June 20, 1686    
                    {\it The Correspondence of Isaac Newton} ref. \cite{turnbull} p. 438

\bibitem{cotes1}    I. Newton, {\it Principia} ref. \cite{cohen1}p.392

\bibitem{herivel}   J. Herivel, {\it The Backrground to Newton's Principia}
                    ( Oxford at the Clarendon Press, 1965) p. 130      

\bibitem{westfall}  R.S. Westfall, {\it Never at Rest: A Biography of
                    Isaac Newton} (Cambridge University Press, Cambridge
                    1995) p. 383

\bibitem{p304}      Letter from Hooke to Newton, Dec. 9, 1679
                    {\it The Correspondence of Isaac Newton} ref. \cite{turnbull} p. 304

\bibitem{herivel1}   J. Herivel, {\it The Backrground to Newton's Principia}
                    ref. \cite{herivel} pp. 192-198

\bibitem{p309}      Letter from Hooke to Newton, Jan. 6, 1680,
                    {\it The Correspondence of Isaac Newton} ref. \cite{turnbull} p. 309

\bibitem{p446}      Letter from Newton to Halley, July 27, 1686,
                    {\it The Correspondence of Isaac Newton} ref. \cite{turnbull} p. 446

\bibitem{p307}      The diagram in this letter has often been reproduced incorrectly.
                    {\it The Correspondence of Isaac Newton} ref. \cite{turnbull} p. 307

\bibitem{michael}   M. Nauenberg, {\it Newton's Early Computational Method for Dynamics},
                    Archive for History of Exact Sciences, {\bf 46} (1994) 221-251

\bibitem{hooke3}    Letter from Hooke to Newton, January 6, 1679/1680
                    {\it The Correspondence of Isaac Newton} ref. \cite{turnbull} p. 309

\bibitem{whiteside2} {\it The Mathematical Papers of Isaac Newton} vol. VI 1684-1691, ed.
                     D.T. Whiteside (Cambridge University Press, 1974) pp. 89-91
                   

\bibitem{whiteside3}  D. T. Whiteside (private communication, 2005)

\bibitem{westfall2}  Quoted in ref. \cite{westfall}  p. 403

\end{thebibliography}
\end{document}